\documentclass[prd,onecolumn,aps,nofootinbib,eqsecnum,notitlepage,nobibnotes,superscriptaddress,preprintnumbers,floatfix]{revtex4-1}

\usepackage[utf8]{inputenc}
\usepackage{color}
\usepackage{amsfonts}
\usepackage{graphicx}
\usepackage[dvipsnames]{xcolor}
\definecolor{refs}{RGB}{245,156,74}
 \usepackage[colorlinks=true,hyperfootnotes=true,citecolor=cyan]{hyperref}
 \usepackage{appendix}
 
\usepackage{dsfont} 
\usepackage[euler]{textgreek}
\usepackage{enumerate}
\usepackage{soul}

\pdfoutput=1

\usepackage{amsmath,amsthm,amssymb}
\usepackage{bm}
\usepackage{soul}
\usepackage{physics}
\DeclareMathOperator{\arcsinh}{arcsinh}

\newcommand{\bi}{\begin{itemize}}
\newcommand{\ei}{\end{itemize}}
\newcommand{\ben}{\begin{enumerate}}
\newcommand{\een}{\end{enumerate}}

\newcommand{\bwt}{\begin{widetext}}
\newcommand{\ewt}{\end{widetext}}

\newcommand{\be}{\begin{equation}}
\newcommand{\ee}{\end{equation}}
\newcommand{\bea}{\begin{eqnarray}}
\newcommand{\eea}{\end{eqnarray}}
\newcommand{\ba}{\begin{array}}
\newcommand{\ea}{\end{array}}

\def\lsim{\mathrel{\rlap{\lower4pt\hbox{\hskip1pt$\sim$}}
    \raise1pt\hbox{$<$}}}                
\def\gsim{\mathrel{\rlap{\lower4pt\hbox{\hskip1pt$\sim$}}
    \raise1pt\hbox{$>$}}}                

\usepackage{tasks}
\begin{document}
\title{Particle Production in Accelerated Thin Bubbles}

\author{\ Florencia Anabella Teppa Pannia}
\email{f.a.teppa.pannia@usal.es}
\affiliation{Departamento de F\'{\i}sica Fundamental and IUFFyM, Universidad de Salamanca, E-37008
Salamanca, Spain.}
\author{Santiago Esteban Perez Bergliaffa}
\email{sepbergliaffa@gmail.com}
\affiliation{Departamento de F\'{\i}sica Te\'{o}rica,
Instituto de F\'{\i}sica, Universidade do Estado de Rio de Janeiro, 20550-013, Rio de Janeiro, Brazil.}
\author{Nelson Pinto-Neto}
\email{nelsonpn@cbpf.br}
\affiliation{CBPF\,-\,Centro Brasileiro de Pesquisas F\'{\i}sicas, rua Xavier Sigaud 150, 22290-180, Rio de Janeiro, Brazil.}

\begin{abstract}
We investigate the creation of scalar particles inside a region delimited by a bubble which is expanding with non-zero acceleration.
The bubble is modelled as a thin shell and plays the role of a moving boundary, thus influencing the fluctuations of the test scalar field inside it. 
Bubbles expanding in Minkowski spacetime as well as those  
dividing two de Sitter spacetimes are explored in a unified way. 
Our results for the Bogoliubov coefficient $\beta_k$  in the adiabatic approximation show that in all cases the creation of scalar particles  
decreases with the mass, and is much more significant in the case of nonzero curvature.  
They also show that the dynamics of the bubble and its size are relevant for particle creation, but in the dS-dS case the combination of both effects leads to a behaviour different from that of Minkowski space-time,
due to the presence of a length scale (the Hubble radius of the internal geometry).

\end{abstract}

\maketitle

\section{Introduction}
\label{sec:introduction}
The propagation of de Sitter bubbles in a given background is relevant in several contexts.
In eternal inflation \cite{Guth2007},
bubbles of new vacuum
nucleate into an ambient region, leading to inflationary patches in different environments.
During their evolution, bubbles may collide and leave an imprint in the cosmic microwave background \cite{Aguirre2009}. Since bubbles may also nucleate in
matter or radiation backgrounds \cite{Simon2009}, their propagation was studied
for homogeneous and inhomogeneous (dust-filled) environments in
\cite{Fischler2008, Simon2009, Casadio2011},
and radiation-filled ambiences
\cite{Ertan2008, TeppaPannia2017}. In particular, the propagation
in inhomogeneous backgrounds can be taken as a first step in addressing the problem of
the generality of inflation
(see for instance \cite{Goldwirth1993,Goldwirth1989b,Goldwirth1990,East2016}).

Another problem of interest in this context is the possible influence of the details of the motion of the wall on the physical phenomena inside it. 
 The impact of a moving spherical cavity on the vacuum in flat spacetime was studied in
\cite{Liberati1998, Mazzitelli2006} (see also
\cite{Dalvit2006}). Such an influence is to be expected, since the bubble can be thought of as a Casimir cavity with a moving boundary, as analysed for instance in \cite{McInnes2008, Milton2012}. 
 We shall examine here the effects of such motion on the vacuum fluctuations of a test scalar field inside an inflating region. 
This problem can be considered as a first step in the study of a complete inflationary model, in which the spectrum of the fluctuations of the inflaton (which yield the  seeds of the structure we observe today) may be affected by the details of the expansion of the region that smoothly interpolates between the expanding patch and its surroundings.  
 
Specifically, the aim of our work is to study the effect of the features of the motion of the boundary on the creation of particles for a test scalar field in two different types of bubbles, in a unified way.
First, the case of a spherical bubble separating two regions in Minkowski spacetime will be considered (see \cite{Setare2001} for the massless case). In this setup, the bubble is assumed to be at rest in the far past and in the distant future, and it smoothly expands between these two states. 
The second case we shall consider is that of a bubble separating two different de Sitter regions, for which the analytic form for the evolution of the radius of the bubble was given in \cite{Simon2009} (see also \cite{Ng2011} for the case of a bubble separating a de Sitter region from a Schwarzschild-de Sitter exterior spacetime). 
 Here the bubble starts with a given initial radius and zero initial velocity, and expands until its radius reaches the asymptotic limit.

The paper is organised as follows. In Section~\ref{sec:model} we present the general framework for the quantization of the test scalar field, and the equation obeyed by the normal modes, with the spherical bubble described by a time-dependent radius $a(t)$. 
We also introduce the vacuum state chosen for the calculations in each case. 
Our results for the creation of particles for specific examples of the evolution of the shell are presented in Section~\ref{sec:examples}, using a choice of the parameters of the radius of the shell that unifies the Minkowski and de Sitter-de Sitter cases. 
We close with our conclusions in Section~\ref{sec:discussion}.

\section{Quantum vacuum fluctuations in an expanding shell}
\label{sec:model}
\subsection{General Framework}
\label{subsec:framework}
Let us consider the evolution of a minimally coupled test scalar field $\Phi$ propagating in a spherically-symmetric de Sitter patch embedded in a generic external background.  As we shall see below, this scenario encompasses also that of a bubble separating two regions of Minkowski spacetime. 
 As mentioned before, the transition between the inner and the outer regions can be modelled by an expanding shell using the thin-shell formalism (see for instance \cite{Israel1966}). 
It follows from the application of such formalism to the case at hand that the time dependence of the radius of 
the bubble is influenced by the internal and external geometries (see for instance
\cite{Fischler2008,TeppaPannia2017}). 
We will describe the inflating region as
a spherically-symmetric bubble with a time-dependent radius $a(t)$ and surface energy density $\sigma$.

The Lagrangian density of the scalar field, in units $c=\hbar=1$, is 
\be
{\cal L}=\frac{1}{2}\sqrt{-g}\left\{g^{\mu\nu}\partial_\mu \Phi \partial_\nu \Phi- {m}^2 \Phi^2 \right\}\,,
\ee
where ${m}^2$ is the mass of the test scalar field. 
 Setting the variation of the action
$S=\int{{\cal L} {\rm d}^4 x}$ 
w.r.t. $\Phi$ equal to zero yields the equation
\be
\label{field_eq}
\left(\Box +{m}^2\right)\Phi=0\,,
\ee
where $\Box \equiv g^{\mu\nu} \nabla_\mu\nabla_\nu = \sqrt{-g}\partial_\mu[\sqrt{-g}g^{\mu\nu}\partial_\nu ]$. 
We consider the de Sitter geometry with line element expressed as
\be
\label{metric:deSitter}
{\rm d}s^2=C^2(\eta)\left({\rm d}\eta^2-{\rm d}\vec{x}^2\right)\,,
\ee
where $C(\eta)=-R/\eta$,  with {$R= 1/H= 3/\sqrt{\Lambda}$},
and the conformal time $\eta$ is defined as 
${\rm d}\eta=e^{-t/R}{\rm d}t$, with the interval $t\in[0,\infty)$ mapped onto $\eta\in[-1,0)$. In the case of the Minkowski geometry, we shall set $C=1$ ($\eta\equiv t$).

Implementing the field transformation \straightphi$(\eta)=C(\eta){\Phi}(\eta)$, eq.~\eqref{field_eq} becomes
\be
\label{field_eq_eta}
\ddot{{\textrm \straightphi}} + \left(C^2 {m}^2 - \frac{\ddot{C}}{C} \right) {\textrm \straightphi} - \nabla^2 {\textrm \straightphi} = 0\,,
\ee
where the dots represent derivatives w.r.t. the conformal time $\eta$. 
The field \straightphi\, can be quantised following the canonical quantisation scheme, with an appropriate definition of the vacuum state. The corresponding expansion is
\be
\label{creation}
{\textrm \straightphi} = \sum_\alpha\left[\hat{b}_\alpha {\textrm \straightphi}_\alpha+\hat{b}_\alpha^{\dag}{\textrm \straightphi}_\alpha^*\right]\,,
\ee
where the collective index $\alpha$ denotes the set of quantum numbers specifying each mode  ${\textrm \straightphi}_\alpha$. Following the approach presented in \cite{Setare2001,Mazzitelli2006},
we assume that the eigenfunctions ${\textrm \straightphi}_\alpha$ can be expanded in terms of an instantaneous basis $\bar{\phi}_\alpha(\vec{x};\bar{a})$, 
where $\bar{a}$ is the instantaneous physical radius of the bubble, in such a way that ${\cal Q}_{\alpha\beta}(\eta) \bar{\phi}_\beta(\vec{x};\bar{a})$
 would be the corresponding eigenfunctions for a static sphere with constant radius $\bar{a}$ and time dependence given by ${\cal Q}_{\alpha\beta}(\eta)$. 
Then, considering $\bar{a} \rightarrow a(\eta)$ and $\bar{\phi}_\alpha \rightarrow \phi_{\alpha}({\vec{x},a(\eta))}$, the field modes are
\be
\label{modes_ansatz}
{\textrm \straightphi}_\alpha(\vec{x},\eta)=\sum_\beta{q_{\alpha\beta}(\eta)\phi_\beta(\vec{x},a(\eta))}\,,
\ee
where the elements of the instantaneous basis are expressed at each moment in spherical coordinates as
\be
\label{static_phi}
\phi_\beta(\vec{x},a(\eta))=\frac{\sqrt{2}\,j_l(j_{l,n}r/a(\eta))}{\left(a(\eta)\right)^{3/2}j'_l(j_{l,n})}Y_{l}^m(\theta,\varphi)\,,
\ee
with $\beta=(l,m,n)$ the set of quantum numbers $(l=0,1,2,\ldots$; $-l\leq m\leq l$; $n=1,2,\ldots)$,  $Y_{l}^m(\theta,\varphi)$ the spherical harmonic functions, $j_l$ the spherical Bessel functions of the first kind, and $j_{l,n}$ their $n$-th zero.\footnote{Bessel's functions of the second kind are discarded under the assumption of regularity at the origin.} Note that the set of quantum numbers is the same for the static and the dynamic cases.
 The basis of eigenfunctions (\ref{static_phi}) satisfies the orthonormality relation
\be
\label{ortho_phi}
\int{\phi_\beta\phi_{\beta'}^\ast{\rm d}^3x}=\delta_{\beta\beta'}\,,
\ee
with Dirichlet's boundary conditions imposed on the hyper-surface $\Sigma$
with time-dependent radius $r = a(\eta)$, that is,
\be
\phi_\beta|_{r=a(\eta)}=0\,.
\ee
Such boundary condition is compatible 
with the assumption 
that no 
transfer of matter or momentum through the thin-shell occurs during the evolution. 
Replacing the \emph{Ansatz} \eqref{modes_ansatz} into the field equation \eqref{field_eq_eta}, we obtain
\be
\label{eq:phiq}
\sum_\beta \left\{\phi_\beta\left[\ddot{q}_{\alpha\beta}+\left(w^2_{\beta} + {m}^2  C^2-\frac{\ddot{C}}{C} \right) q_{\alpha\beta}\right]  +  2 \dot{q}_{\alpha\beta} \dot{\phi}_{\beta} + q_{\alpha\beta} \ddot{\phi}_\beta\right\} =0\,,
\ee
where $w^2_{\beta}=\left(j_{l,n}/a(\eta)\right)^2$ are the time-dependent eigenfrequencies.
 This expression can be simplified by multiplying by $\phi^\ast_{\beta'}$ and integrating over the region inside the sphere at a given moment $\eta$, thus deriving
 \cite{Setare2001}
\be
\ddot{q}_{\alpha\beta}+\left[w^2_{\beta} + {m}^2 C^2  -\frac{\ddot{C}}{C} \right] q_{\alpha\beta}= 
\sum_{\beta'}\left(2\dot{q}_{\alpha\beta'}g^{(1)}_{\beta\beta'} +{q}_{\alpha\beta}g^{(2)}_{\beta\beta'}\right)\,,
 \ee
 where the coefficients on the r.h.s. are defined as follows:
 \begin{eqnarray}
 \label{coeff_gint}
g^{(1)}_{\beta\beta'}&=&-\int{\dot{\phi}_{\beta'}\phi^{\ast}_\beta {\rm d}^3x}\,, \\ 
\label{coeff_gint2}
g^{(2)}_{\beta\beta'}&=& -\int{\ddot{\phi}_{\beta'}\phi^{\ast}_\beta {\rm d}^3x}\,, 
\end{eqnarray}
with the integrations defined over the region inside the sphere of radius $a(\eta)$, and
 $ g^{(2)}_{\beta\beta'} =\dot{g}^{(1)}_{\beta\beta'} +\sum_\gamma{ g^{(1)}_{\beta'\gamma}g^{\ast(1)}_{\beta\gamma}}.$
This last relation between the coefficients can be found by taking into account the completeness condition for the eigenfunctions, $\sum_\gamma{\phi_\gamma(\vec{x})\phi^\ast_\gamma(\vec{x}')=\delta(\vec{x}-\vec{x}')}$. Furthermore,
the standard orthonormality relations for the spherical harmonics, together with the following integral involving the spherical Bessel functions
\be
\int^1_ 0{z^2j_l(bz)j_l(cz)}{\rm d}z= \frac{bj'_l(b)j_l(c)-cj_l(b)j'_l(c)}{c^2-b^2}\,,
\ee
can be used to evaluate the r.h.s. of eqs.~\eqref{coeff_gint} and \eqref{coeff_gint2}, yielding the following expressions for the coefficients: 
\bea
&&g^{(1)}_{lnn'}=hd_{lnn'}\,,  \\
&&g^{(2)}_{lnn'}=\dot{h}d_{lnn'}+h^2\sum_p{d_{lnp}d_{ln'p}}\,, 
\eea
where $h\equiv \dot{a}/a$ is the expansion rate of the bubble radius and
\be
\left\{
\begin{tabular}{rcl}
    $d_{lnn}$&$=$&$0\,,$ \\
    $d_{lnn'}$&$=$&$\frac{2j_{l,n}j_{l,n'}}{\left(j^2_{l,n}-j^2_{l,n'}\right)}\,,$  \hspace{.3cm}  {\rm for}\ $n\ne n'.$
    \end{tabular}
\right. \nonumber
\ee
{Since the coefficients $g^{(1)}$ and $g^{(2)}$ 
are anti-symmetric, it follows that} $q_{\alpha\beta}$ is independent of the quantum number $m$, and can be chosen as diagonal w.r.t. $l$ and $n$ as 
$q_{\beta\beta'}=\delta_{l l'}\delta_{mm'}q_{lnn'}$ \cite{Setare2001}. 
 This leads to the following infinite set of coupled differential equations for the coefficients $q_{lnn'}$:
\begin{equation}
 \label{eq:q}
 \ddot{q}_{lnn'}+\left[w^2_{ln'} + {m}^2 C^2 - \frac{\ddot{C}}{C} \right] q_{lnn'}= 
 2h\sum^\infty_{p=1}\dot{q}_{lnp}d_{lpn'} 
 +\dot{h}\sum^\infty_{p=1}{q}_{lnp}d_{lpn'} 
 +h^2\sum^\infty_{p,s=1}{q}_{lnp}d_{lps}d_{ln's}. 
 \end{equation}
 
At this point different types of effects, associated to specific terms in eq.~\eqref{eq:q}, that act on the evolution of the field modes can be specified:
\begin{enumerate}[i)]
\item the time-dependence of the eigenfrequencies $\omega_{ln'}(\eta)= j_{l,n'}/a(\eta)$, which originates from the dependence with time of the radius of the sphere;
\item  the expansion of the de Sitter spacetime, which manifests itself in the terms involving the function $C(\eta)$ and its derivatives (such terms are null in the case of a Minkowskian geometry); and
\item the expansion rate of the bubble radius, through $h$ and $\dot{h}$ (r.h.s. of eq.~\eqref{eq:q}).
  \end{enumerate}
As a first approach to the calculation of the number density of created particles, we will work within an {\it adiabatic approximation} \cite{Setare2001}, that is, neglecting the third contribution. This assumption is reasonable provided that the evolution of the modes is influenced but not dominated by the bubble expansion.
Then, the motion of the bubble 
is only present in the equation for the field fluctuations {through}  $\omega_{ln'}(\eta) = j_{l,n'}/a(\eta)$. The essential advantage of this approximation is that it leads to the decoupling of
the differential equations for each set of quantum numbers $(l,n')$. 
Necessary conditions for this approximation to be valid are 
$\dot a^2\ll j_{l,n'}^2$ and 
$a\ddot a\ll j_{l,n'}^2$
\cite{Setare2001},
which have been numerically verified in the particular examples analysed in the next Section.

The system of equations \eqref{eq:q} is then reduced to
\be
\label{eq:q2}
\ddot{q}_{k}(\eta)+\left[\frac{k^2}{a^2(\eta)} + {m}^2 C^2(\eta) - \frac{\ddot{C}(\eta)}{C(\eta)} \right] q_{k}(\eta)=0\,,
\ee
where, for simplicity, we defined $k\equiv j_{l,n'}$ { and denoted the corresponding functions with the subscript $k$}. The ${q_k}$ are a discrete set of solutions with frequencies given by the zeros of the Bessel functions which, in addition, impose a natural IR cut-off on the range of $k$ for the computation of the particle production: $j_{0,1}=\pi\leq k$. It is worth noting that the fact that the mode frequencies are discrete is a direct consequence of the assumption of Dirichlet boundary conditions at the moving radius (similar results were found in \cite{Mazzitelli2006}).

\subsection{Vacuum choice and initial conditions}
\label{subs:choice}
The evolution of the modes of the fluctuations of the test scalar field inside a given bubble is governed by eq.~\eqref{eq:q2} {and} supplemented by appropriate initial conditions, dictated {in turn} by a vacuum choice. 
 Let us begin with the Minkowski case, for which 
eq.~\eqref{eq:q2} reads
\be
\label{eq:qM}
\frac{{\rm d}^2 q_k(t)}{{\rm d}t^2}+\left[\frac{k^2}{a^2(t)} + m^2\right] q_{k}(t)=0\,,
\ee
where $t$ is Minkowski coordinate time. A
convenient form for the time dependence of the radius of the bubble is the following
\cite{BirrellDavies}:
\be
\label{a-Setare}
a(t) = \frac{1}{\sqrt{A + B\tanh(t/t_0)}}, \qquad A>|B|\,,
\ee
where $A$ and $B$ are constants, and $B$ will be assumed to be negative to represent an expanding bubble. Hence, at $t\rightarrow\mp\infty$, $a(t)$ asymptotically approaches the constant values $a_{{\rm (i,f)}}\equiv(A\pm B)^{-1/2}$. This {\it Ansatz} yields
an analytical solution of eq.~\eqref{eq:qM} for the massless case  and, as  will be seen later, suitable choices of $A$ and $B$ allow the comparison with the exact solution of $a(t)$ for the de Sitter geometry.  
Note that the time scale in eq.~\eqref{a-Setare} is given by the arbitrary parameter $t_0$ (which, as we shall see below, plays the role of $1/H$ in the de Sitter-de Sitter case). Hence, the following dimensionless parameters can be defined: $\bar{t}=t/t_0$, $\bar{m}=m\,t_0$ and $\bar{a}=a/t_0$. These, {in turn,} induce the definition of $\bar{A}=A\,t_0^2$, $\bar{B}=B\,t_0^2$ and $\bar{q_k}=q_k/\sqrt{t_0}$, which are also dimensionless. As eqs.~\eqref{eq:qM} and \eqref{a-Setare} 
are unaltered upon substitution by the barred parameters, particle production will depend only of $\bar{A}$, $\bar{B}$ and $\bar{m}$, and not of $t_0$. This is in accordance with the exact results presented in Ref.~\cite{Setare2001} for the massless case (see eq.~\eqref{exact} below), where $t_0$ is absent when the barred variables are used. In order to simplify the notation, the bars will be omitted from now on.

Due to the nature of the Minkowski background, one can define the ``in'' and ``out" vacua as usual for very early and late times:
\be
\label{vacuum-flat}
q_{k}^{{\rm vac} (i,f)}(t) = \sqrt{\frac{1}{2\nu^{\rm (i,f)}_k}}\exp(-i\nu^{\rm (i,f)}_kt)\,,
\ee
where $\nu^{\rm (i,f)}_k\equiv \sqrt{(k/a_{\rm (i,f)})^2+m^2}$ and $a_{\rm (i,f)}$ is expressed in terms of the parameters $A$ and $B$. This prescription assures a Poincar\'e invariant quantum state \cite{BirrellDavies}. 

The second case that will be examined in this work corresponds to a bubble which is the boundary between an inner de Sitter spacetime embedded into another de Sitter geometry.  This two regions are characterised by cosmological constants $\Lambda$ and $\hat{\Lambda}$, respectively.
The bubble profiles are described by an exact expression
for the radius, which was obtained in Ref.~\cite{Simon2009} using the thin-shell formalism (see also \cite{Takamizu2015}).
 The evolution of the radius of the shell is expressed in the inner coordinate system as \cite{Simon2009}
\begin{equation}
a(t)=\sqrt{{u}^{-2} + [\exp(-tH)-1]^2 }\,,
\label{radiusdsds}
\end{equation}
where $t$ is the cosmic time. The parameter $u^2$ characterises the dynamics of the bubble and is expressed in terms of the parameters of the configuration matched by the thin shell as
\begin{equation}
u^2 \equiv 
\frac{1}{\bar{H}^2}
\left[\frac{\Delta\bar{\Lambda}}{3{\bar{\sigma}}}
+ \frac{{\bar{\sigma}}}{4} \right]^2\,,
\end{equation}
where  
$\Delta\bar\Lambda= (\hat{\Lambda}-\Lambda)/M_{\rm pl}^2$  is the dimensionless difference between the cosmological constants of the outer and inner geometries, 
  $\bar{H}=\sqrt{\bar{\Lambda}/3}$, with 
   $\bar\Lambda\equiv\Lambda/M_{\rm pl}^2$ the dimensionless cosmological constant in the inner region, 
   and $\bar{\sigma}\equiv\sigma/M_{\rm pl}^3$ the dimensionless energy density of the shell,
which must obey ${\bar{\sigma}}<2\sqrt{\Delta\bar{\Lambda}}/\sqrt{3}$ \cite{Simon2009}. 
Hence, the minimum value for $u^2$ corresponds to 
$u_{min}^2\approx \Delta\bar{\Lambda}/\bar{\Lambda}$  and, 
 for small values of $\bar{\sigma}$, we have  
$u^2 \bar{H}^2\approx
\left(\Delta\bar{\Lambda}/(3\bar{\sigma})\right)^2$,
which shows that it is the difference of the cosmological constants that sets the value of $u^2$ for a given $\bar{\sigma}$, 
both for large and small values of $u^2$. 
The bubble described by eq.~\eqref{radiusdsds} is spontaneously created at {$t_i=0$} with null velocity $(\dot a(0)=0)$ and initial radius $a_i\equiv a(0)=1/\sqrt{u^2}$, and asymptotically reaches its maximum value $a_f$ at $t_f\rightarrow \infty$, with $a_f\equiv a(t_f)=\sqrt{1/u^2 + 1}$. To summarise, $a_f^2=a_i^2 + 1$ with 
\bea
\label{profile-DS}
a_i &=& \frac{1}{\sqrt{u^2}}\,, \\
a_f &=& \sqrt{\frac{1}{u^2}+1}\,.
\eea

Let us rewrite eq.~\eqref{eq:q2} using the dimensionless variables $\bar{a}\equiv aH$, $\bar{\eta}\equiv \eta H$ and $\bar{m}\equiv mH$.  
 Removing the bars for simplicity, it reads
\be
\label{eq:q3}
\ddot{q}_{k}(\eta)+\left[\frac{k^2}{\tilde{a}^2(\eta)} - \frac{(2-{m}^2)}{\eta^2}\right] q_{k}(\eta)=0\,,
\ee
with 
\be
\label{aeta}
\tilde{a}(\eta)=\sqrt{u^{-2}+(\eta+1)^2}.
\ee
Since all parameters are in
units of $H$, it is seen from eqs.~\eqref{radiusdsds} and \eqref{eq:q3}  that particle production will depend only on $u^2$ and $m$ in this case. Note that $q_k$ still has dimensions of $[{\rm{length}}]^{1/2}$ and, as we shall see below, there is an upper limit for the mass, given by $m=3/2$.

We are considering bubbles which nucleate with a finite radius and expand up to a final constant radius. The vacuum Bunch-Davies solution
will be chosen as initial quantum state, since it is the unique quantum state which is invariant under the de Sitter group $SO(1,4)$ transformations for $m\neq 0$ that can be extended to the $m=0$ case, being in the latter case invariant under the maximal subgroup $E(3)$ of $SO(1,4)$ \cite{Allen1985}. Hence, the initial conditions are \cite{Riotto:2002yw}:
\begin{eqnarray}
\label{etai}
q_k(\eta_i) &=& \biggl[\frac{\sqrt{\pi}}{2} \sqrt{-\eta}{\cal H}_p^{\rm(1)}\left(-\eta{\bar{k}}(\eta)\right)\biggr]_{\eta_i}\,, \\
\label{detai}
\dot{q}_k(\eta_i) &=& \biggl[  \frac{\rm d}{{\rm d}\eta}\biggl( \frac{\sqrt{\pi}}{2} \sqrt{-\eta}{\cal H}_p^{\rm(1)}\left(-\eta{\bar{k}}(\eta) \right)\biggr)\biggr]_{\eta_i}\,,
\end{eqnarray}
where ${\cal H}_p^{\rm(1)}$ is the Hankel function of the first kind with index $p\equiv\sqrt{9/4-{m}^2}$ (hence, $m^2\leq9/4$), and $\bar{k}(\eta)\equiv k/\tilde a(\eta)$. The solution of eq.~\eqref{eq:q3} satisfying the initial conditions given in eq.~\eqref{etai}-\eqref{detai} will be denoted by  $q_k^{(i)}(\eta)$.
The natural vacuum choice at times beyond some time $\eta_f$ at which the bubble radius becomes comoving with the background is again the Bunch-Davies solution evaluated at $\bar{k}_f\equiv k/\tilde a(\eta_f)$, that is,
\begin{equation}
\label{etaf}
q_k^{(f)}(\eta) = \frac{\sqrt{\pi}}{2}\sqrt{-\eta}{\cal H}_p^{\rm(1)}\left(-{\bar{k}}_f \eta\right)\,.
\end{equation}

\section{Particle creation: Numerical results for particular models}
\label{sec:examples}

In this section, we will calculate numerically some quantities asociated to particle creation in a variety of bubbles expanding in a Minkowski background
according to eq.~\eqref{a-Setare}, and compare the results with those obtained in the corresponding case of the de Sitter-de Sitter bubbles.

The sets of functions that define the ``in'' and ``out'' vacuum states in the Minkowski and the de Sitter case, respectively given by  eq.~\eqref{vacuum-flat} and eqs.~\eqref{etai} and \eqref{etaf}, are a basis of the solution space satisfying the normalisation condition
\be
i\left(q_k^{(i)*} \dot{q}^{(f)}_k- q_k^{(i)*} \dot{q}^{(f)}_k\right)=1\,.
\ee
Hence, one can write the expansion
\be
q_k^{(f)}(\eta)=\alpha_k^{(i,f)} q_k^{(i)}(\eta) + \beta_k^{(i,f)} q_k^{(i)\ast} (\eta)\,,
\ee
where
\bea
\alpha_k^{(i,f)} &=&i(q_k^{(i)*} \dot{q}^{(f)}_k- \dot{q}_k^{(i)*} q^{(f)}_k)\,, \\
\beta_k^{(i,f)} &=&-i(q_k^{(i)} \dot{q}^{(f)}_k- \dot{q}_k^{(i)} q^{(f)}_k)\,,
\eea
are the so-called Bogoliubov coefficients \cite{BirrellDavies}.
The number density of particles per frequency associated with the $\ket{(i)}$ state at the time $t_f$ at which the ``out'' vacuum is defined 
is then given by
\be
\label{eq:nk0}
n_k^{(i,f)} \equiv \bra{i}\hat{N}_k^{(f)}\ket{i} = \bra{i}\hat{b}_k^{\dag (f)}\hat{b}_k^{(f)}\ket{i}\,,
\ee
yielding (see Ref.~\cite{BirrellDavies})
\be
\label{eq:nk}
n_k^{(i,f)} = \left|\beta_k^{(i,f)}\right|^2 =\left|q_k^{(i)} \dot{q}^{(f)}_k - \dot{q}_k^{(i)} q^{(f)}_k\right|_
{t=t_f}^2\,.
\ee
Reintroducing the original indices, $k\rightarrow (l,n)$, the total number of particles  reads
\be
\label{eq:n}
N = \sum_{l=0}^{\infty}(2l+1) \sum_{n=1}^{\infty}n_{ln}^{(i,f)}\,.
\ee

\begin{figure*}
\resizebox{8.7cm}{!}{\includegraphics[angle=0]{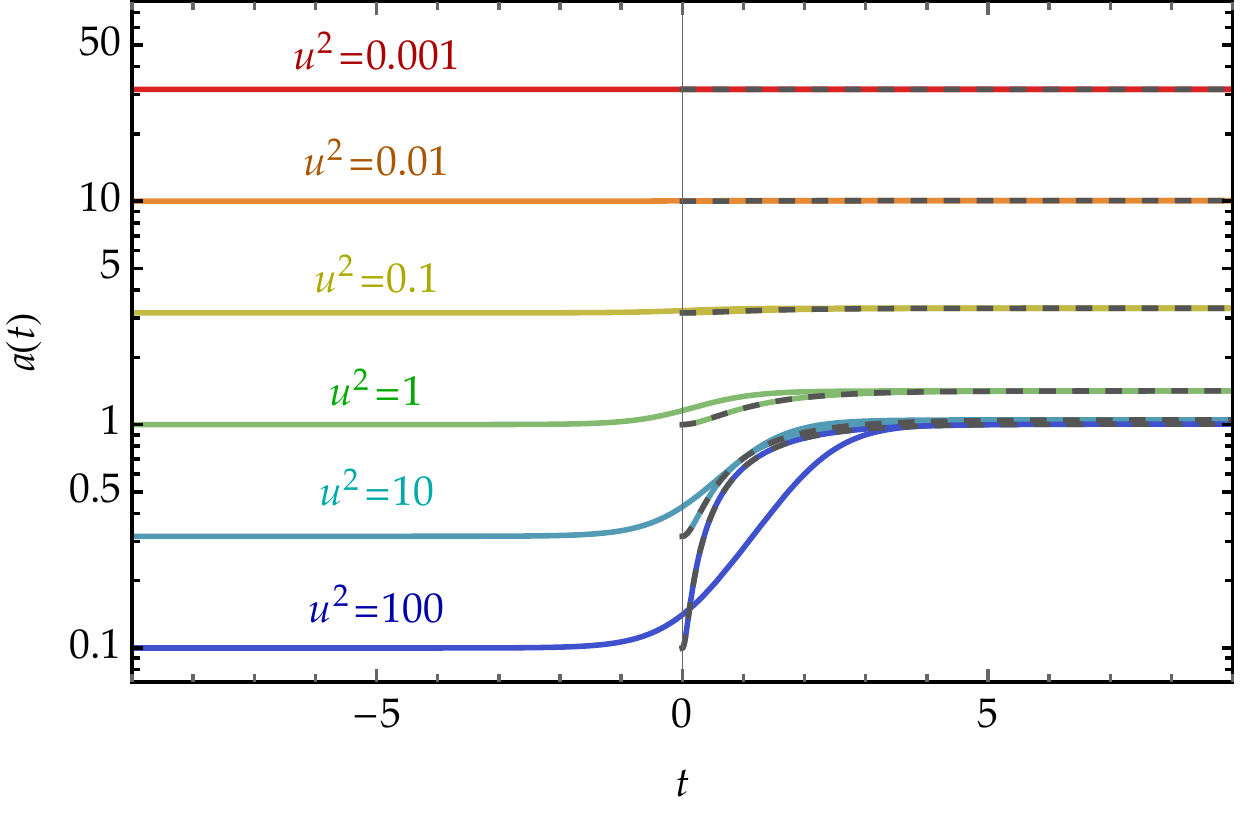}}
\hspace{0.2cm}
\resizebox{8.7cm}{!}{\includegraphics[angle=0]{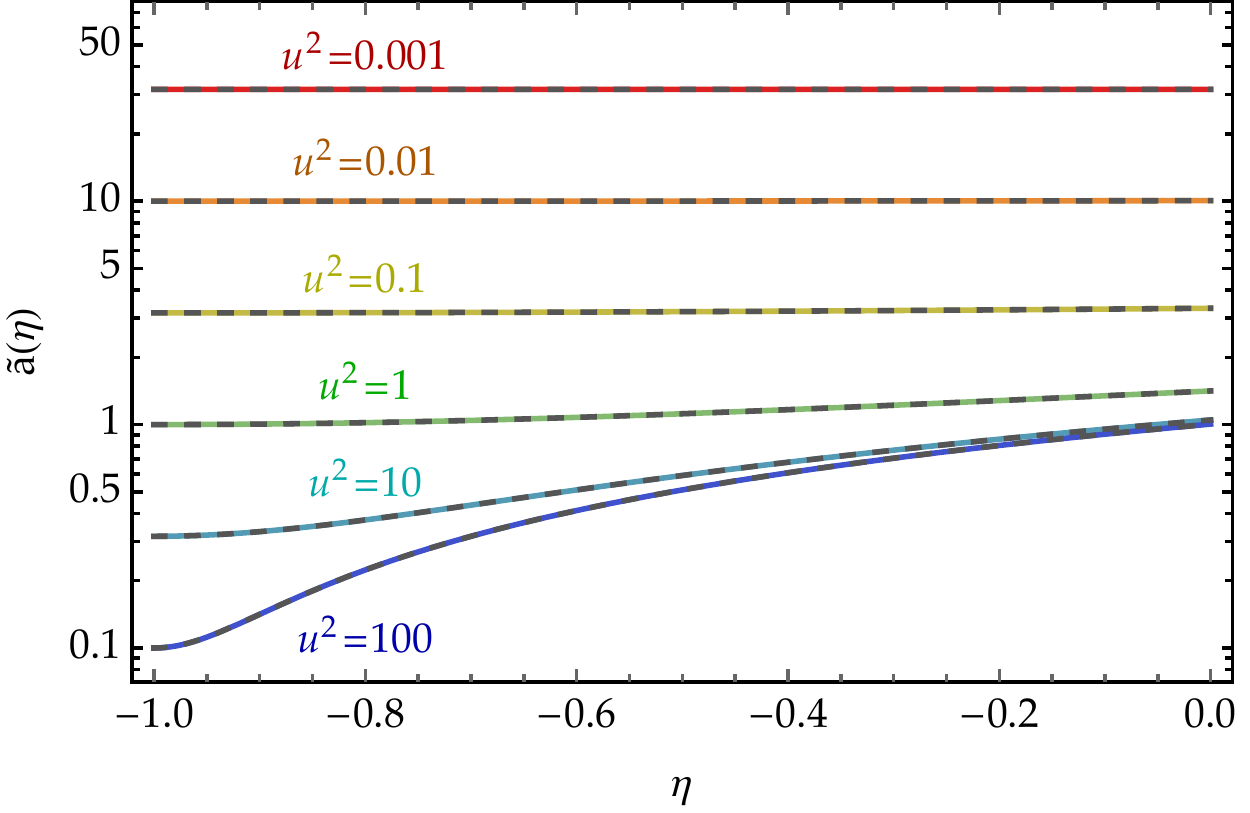}}
 \caption{The figure shows the evolution of the radius of the expanding bubble for different representative values of the parameter $u^2$, as a function of coordinate and conformal time, $t$ and $\eta$ (left and right panels, respectively). Solid coloured lines represent the {\it Ansatz} for $a(t)$, given by eq.~\eqref{a-Setare}, used in the Minkoswki case. Dashed gray-coloured curves indicate the corresponding radius of the shell for the de Sitter case, expressed as a function of the coordinate and conformal times by eqs.~\eqref{radiusdsds} and \eqref{aeta}, respectively. In both cases, the  parameter $u^2$ completely determines the initial and final radius of the bubbles.}
  \label{bubevol}
  \end{figure*}
  
Before presenting the results of the calculation of the particle number, let us introduce a choice of parameters that will lead to a unified description of both cases. 
 We have seen that in the de Sitter case the profile of the expanding bubble depends on one single parameter, $u^2$, while in the Minkowski case it depends on $A$ and $B$. The profiles are extremely constrained in the former case due to the mandatory use of the junction conditions in the matching of the de Sitter spacetimes, which are not necessary in the latter.
  As our aim is to compare the effects of the de Sitter background on particle production, we shall constrain the possible bubble evolutions in the Minkowski case by choosing $a_i^{\rm M}=a_i^{\rm DS}$, $a_f^{\rm M}=a_f^{\rm DS}$, the indices M and DS referring to the Minkowski and de Sitter cases, respectively. Note, however, that $t_{in}^{\rm M}\rightarrow -\infty$ and $t_{in}^{\rm DS}=0$ will be kept. 
Such a choice implies that $(A+|B|)^{-1/2}=1/\sqrt{u^2}$
and $(A-|B|)^{-1/2}=(1/u^2+1)^{1/2}$
 Hence, each choice of {$u^2$} determines the parameters $A,B$ in the Minkowski case through
\begin{eqnarray}
\label{ABu1}
A&=& \frac{u^4+2u^2}{2(u^2+1)\,}, \\
\label{ABu2}
|B|&=& \frac{u^4}{2(u^2+1)}\,.
\end{eqnarray}
In this way, particle creation in the Minkowski bubble will be compared with the same process in the de Sitter bubble by making different choices of the parameters $u^2$ in both cases, with $A$ and $|B|$ given by eqs.~\eqref{ABu1} and \eqref{ABu2} in the Minkowski case. 
 This parameter $u^2$ is a measure of
 the initial and final radius of a given bubble, and also of the relative size of the bubbles: larger $u^2$ means
 smaller initial and final radius, and larger difference between them. The latter means  
 larger velocity (hence larger acceleration and deceleration 
 while the bubble joins the initial and final state). These features  are displayed in Fig.~\ref{bubevol} for representative values of the parameter $u^2$ within the range $10^{-3}\le u^2 \le 10^2$. 

Regarding the dependence of the field modes with the mass parameter $m$, let us point out that while $t_0$ is arbitrary (due to the absence of a scale in Minkowsky spacetime), $H$ is expected to correspond to very high energies ($\sim 10^9$\,GeV or more). Hence, while the mass range in the Minkowski case is rather arbitrary, in the de Sitter case masses much lower than $H$ are expected for any scalar field. Furthermore, as we have seen, in the latter case the mass is constrained by $m^2\leq 9/4$. In the numerical calculations, the range of values $0\le m\le0.5$ will be considered.

\subsection{Bubble expanding in a Minkowski background} 

\begin{figure*}
  \resizebox{9.cm}{!}{\includegraphics[angle=0]{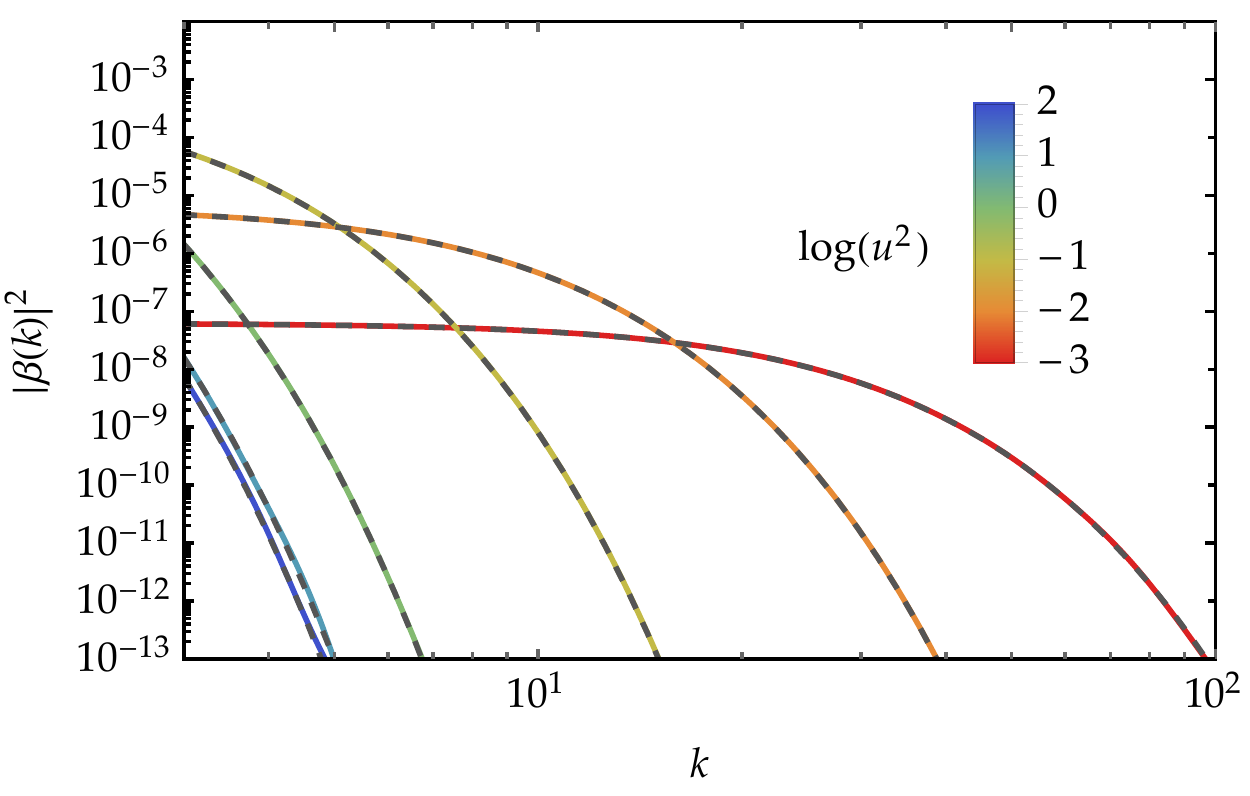}}
  \hspace{0.cm}
  \resizebox{8.7cm}{!}{\includegraphics[angle=0]{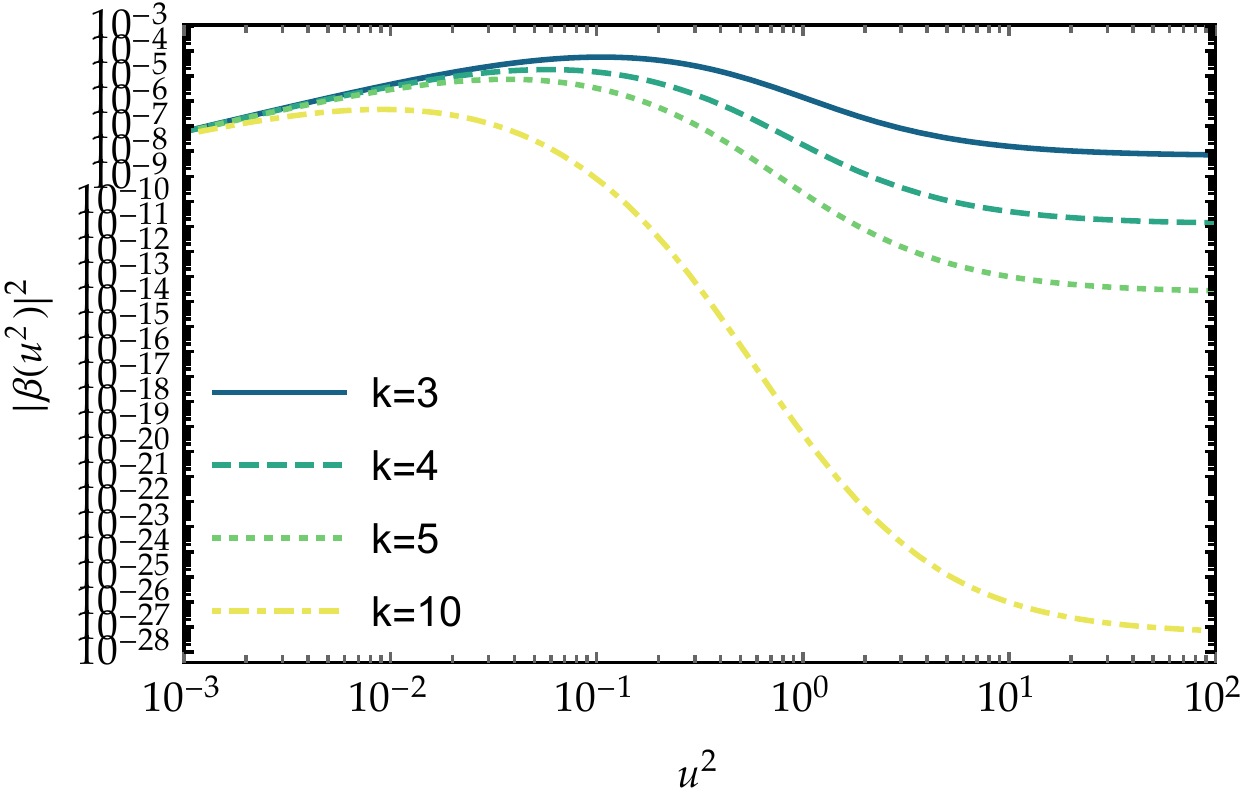}} \\
  \resizebox{8.7cm}{!}{\includegraphics[angle=0]{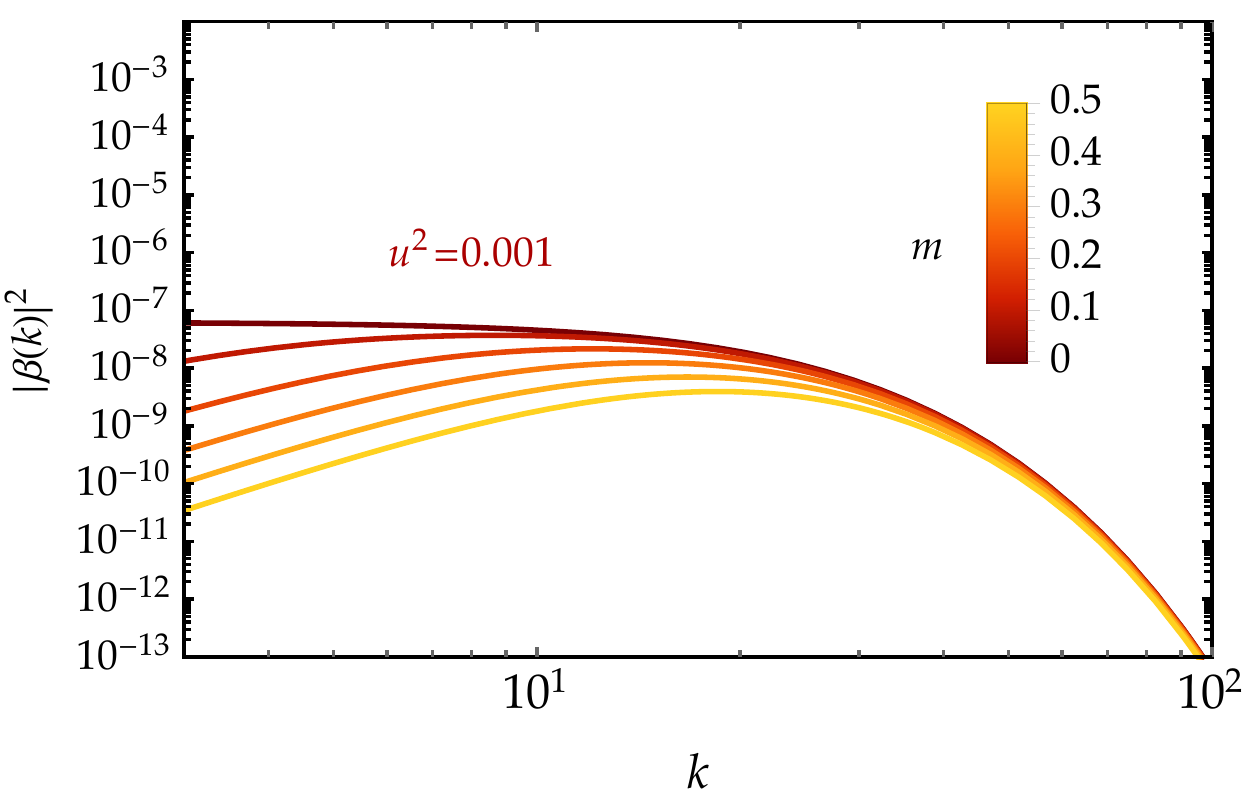}} 
  \hspace{0.3cm}
    \resizebox{8.7cm}{!}{\includegraphics[angle=0]{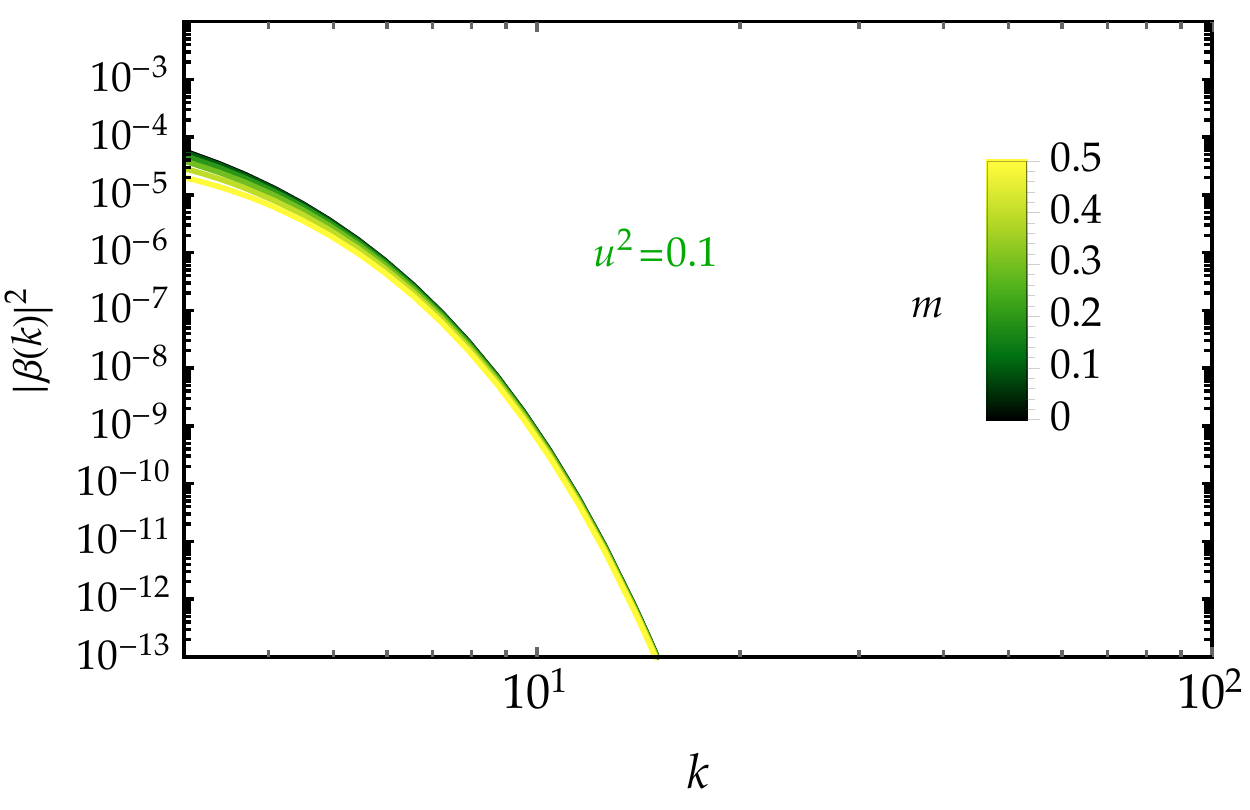}} 
 \caption{The figure shows the number of particles per mode as a function of $k$ and $u^2$ for a bubble expanding in a Minkowski background. The parameter $u^2$, which characterizes the bubble evolution, takes representative values within the range  $10^{-3}\leq u^2\leq 10^2$. On the top left panel are displayed the cases with $m=0$, together with the analytical solution given by eq.~\eqref{exact} and shown in gray-dashed lines. The top right panel shows the existence of a maximum for $|\beta_k|^2$ for low values of $u^2$. 
 The bottom left (right) panel shows $|\beta_k|^2$ as a function of $k$ for $u^2=0.001$  ($u^2=0.1$) and several values of the mass  within the range $0\le m \le 0.5$. 
 As expected, the particle number decreases for larger values of $m$, although the decrement is less
 significant
 for larger values of $u^2$. 
 It is worth noting that in spite of the fact that it is displayed as a continuous variable, only discrete values of $k$ 
 (namely those corresponding to the zeros of the spherically-symmetric Bessel's functions)
 contribute to $|\beta_k|^2$ .}
  \label{fig:beta2MIN}
  \end{figure*}

We first start by checking our numerical approach through the comparison between the analytical and numerical solutions for the massless case. The analytical expression for the number of particles per mode presented in Ref.~\cite{Setare2001} is given by 
\be
\label{exact}
n_k^{(i,f)} =\frac{\sinh^2(\pi \hat{\nu}_{k})}{\sinh (\pi \nu^{\rm (i)}_{k})\sinh(\pi\nu^{\rm (f)}_{k})}\,,
\ee
with $\hat{\nu}_{k}\equiv\frac 1 2 \left(\nu^{\rm (i)}_{k}-\nu^{\rm (f)}_{k}\right)$ and $\nu^{\rm (i,f)}_k = k/a_{\rm (i,f)}$. 
The plot of this expression is displayed in the top panel on the left of Fig.~\ref{fig:beta2MIN} for different values of the parameter $u^2$, together with the results for $|\beta_k|^2$ of our numerical calculation for $m=0$ (based on eqs.~\eqref{eq:qM} and \eqref{vacuum-flat}), showing a very good agreement between them.

The results also suggest that for any small fixed value of $k$, $|\beta_k|^2$ has a maximum as a function of $u^2$, a fact which is confirmed by the plots in the top right panel of Fig.~\ref{fig:beta2MIN}. Such a behaviour may be understood, in the adiabatic approximation used here, as the result of two competing  effects: the dynamics of the bubble and its size. 
For low $u^2$, the dynamics of the bubble becomes more important with increasing $u^2$, while the radius of the bubble decreases.
The dynamics of the bubble dominates the creation of particles up to the point where the bubble is small enough to lower it, until a maximum value is reached (see top right panel of Fig.~\ref{bubevol}). 
For even larger values of $u^2$,  $|\beta_k|^2$ decreases until
particle creation stabilises and becomes independent of $u^2$, in the adiabatic approximation, as shown in Fig.~\ref{bubevol}. 
 Let us consider some relevant limits in Eq.~\eqref{exact} in order to better understand the dependence of $|\beta_k|^2$ on the parameter $u^2$ for the case $m=0$: \\

\paragraph{} For $u^2 \ll 1$, we obtain $\nu_k^{(i)}=ku$, $\nu_k^{(f)}\approx ku\left(1-\frac{u^2}{2}+{\cal O}(u^4)\right)$, and $\hat \nu_k\approx\frac k 4 u^ 3$. 
 It follows  then that 
 \begin{equation}
 n_k^{(i,f)}= \frac{u^4}{4}\left[1-u^2\left(\frac 3 4 +\frac{\pi^2}{3}k^2
 \right)+{\cal O}(u^4)
 \right],
  \end{equation}
 which for small values of $k$ agrees with our numerical results  for the lowest value of $u^2$ in the left top panel of Fig.~\ref{fig:beta2MIN}. Hence, the number of created particles goes to zero for small $u^2$ as expected, and grows with $u$ for small $u$, in accordance with the plots in the first panel of  Fig.~\ref{fig:beta2MIN}. \\
 
 \paragraph{} For $u^2 \gg 1$, on the other hand, we have $\nu_k^{(i)}=ku$, $\nu_k^{(f)}=ku(u^{-1}+{\cal O}(u^{-3}))$, and then $\hat \nu_k\approx \frac k 2 (u-1)$. 
  Hence it follows from Eq.~\eqref{exact} that
  \begin{equation}
  n_k^{(i,f)}\approx \frac{\left[\cosh\left(\frac {\pi}{2}k \right)-
\sinh\left(\frac {\pi}{2}k \right)\right]^2  }{\sinh(\pi k)},
  \label{npa}
  \end{equation}
  which is also in good agreement with the numerical results for the highest value of $u^2$ in the top left panel of Fig.~\ref{fig:beta2MIN}.
Hence, the number of created particles does not depend of $u^2$ for large $u^2$, in the adiabatic approximation. \\

 The above results can be also qualitatively extended to the massive cases, as shown in the bottom panels of Fig.~\ref{fig:beta2MIN}. In these plots, we display the results of the corresponding numerical calculation of $|\beta_k|^2$ for $u^2=0.001$ and $u^2=0.1$, and different values of the mass parameter. As expected in the case at hand, $|\beta_k|^2$ decreases with increasing $m$, for any fixed value of $u^2$, and this spread in $m$ decreases with increasing $u^2$.  

Let us finally remark that only the values of $k$ corresponding to the zeros of the spherical Bessel functions contribute to $|\beta_k|^2$. The plots in Fig.~\ref{fig:beta2MIN} (and in all subsequent plots of $|\beta_k|^2$)
have been constructed taking $k$ as a continuous variable to display in a more appropriate way the dependence of $|\beta_k|^2$ with the parameters $u^2$ and $m$. 
 Then, due to the discrete nature of the spectrum and its IR cut-off mentioned above ($\pi \le k$), the actual total effective particle production is  reduced (specially in the case of large values of $u^2$, for which the main contribution to $|\beta_k|^2$ comes from small values of $k$, where the set of Bessel's zeros is less dense.

\subsection{Bubble expanding in a de Sitter-de Sitter background}

We begin this section by presenting some analytical results to explore particle production within relevant limits and to be used to check the numerical calculations. For such a goal, it is useful to first rewrite eq.~\eqref{eq:q3}
using the transformations
${\rm d}\tau={\rm d}\eta/\tilde{a}(\eta)$ and 
$\chi_k(\tau)=q_k(\eta)/\sqrt{\tilde{a}(\eta)}$, with $\tilde{a}(\eta)$ given by eq.~\eqref{aeta}. 
 The former leads to the relation $\eta+1=(\sinh{\tau})/u$, with the interval $\eta\in[-1,0)$ mapped onto $\tau\in[0,\ln (u+\sqrt{1+u^2}))$. The transformed equation reads 
\begin{equation}
  \label{eqz}
    \frac{{\rm d}^2\chi_k}{{\rm d}\tau^2}+\left[k^2-\frac{(2-m^2)\hat{a}(\tau)^2}{\eta(\tau)^2}-\frac{1}{z(\tau)} \frac{{\rm d}^2z(\tau)}{{\rm d}\tau^2}\right]\chi_k=0\,,
\end{equation}
with $\hat{a}(\tau)=(\cosh{\tau})/u$ and $z(\tau)\equiv 1/\sqrt{\hat{a}(\tau)}$. From the definition of $\hat{a}(\tau)$ it follows that 
$\left|\frac 1 z \frac{{\rm d}^2z}{{\rm d}\tau^2}\right|\leq\frac 1 2\,.$
Hence, such a term (which has a smooth evolution) can be neglected for $k\gg1$. For small values of $\eta$, it is the term $(2-m^2)a^2/\eta^2$ (which grows for $\eta\rightarrow 0$) that governs the evolution. We will consider $0\leq m \leq 1$, hence $1\leq 2-m^2\leq 2$. Finally, eq.~\eqref{eq:q3} can also be written only in terms of $\tau$ as
\begin{equation}
    \frac{{\rm d}^2\chi_k}{{\rm d}\tau^2}+\left[k^2-\frac{(2-m^2)\cosh^2\tau}{(\sinh\tau-u)^2}+\frac 1 2 -\frac 3 4 \tanh^2\tau \right]\chi_k=0\,,
    \label{eqtau}
\end{equation}
within the interval $\tau \in [0,\ln (u+\sqrt{1+u^2}))$.

Let us fix the value of the conformal time at which our numerical calculations stop as $|\eta_f|=10^{-4}$, and consider two particular cases: \\

\paragraph{$\bar k_f |\eta_f| \gg 1$, $0<u<\infty$\,.}   Since 
$\bar{k}_i/\bar{k}_f=\sqrt{1+u^2}\gsim 1\,,$
it follows that
$\bar k_i |\eta_f|\gg 1$. 
In order to calculate $|\beta_k^{(i,f)}|^2$, (see eq.~\eqref{eq:nk}), we have to evaluate $q_k^{(i)}(\eta)$, $q_k^{(f)}(\eta)$, and their first derivatives at $\eta_f$.
The mode $q_k^{(f)}(\eta)$ around $|\eta_f|$ is given by eq.~\eqref{etaf}. As we are assuming $\bar k_f |\eta_f| \gg 1$, then, for $|\eta|=|\eta_f|+\epsilon$, $|\epsilon|\ll1$, one gets the ``out'' vacuum mode as
\begin{equation}
q_k^{(f)}(\eta)= \frac{1}{\sqrt{2\bar k_f}}e^{i\bar k_f\eta}\left(1+O((\bar k_f\eta)^{-1})
\right)\,,
\label{qfetaf}
\end{equation}
where $\bar k_f=k/a_f=ku/\sqrt{u^2+1}$. 
We have now to evaluate $q_k^{(i)}(\eta)$ around $|\eta_f|$, which are solutions of eq.~\eqref{eq:q3} (or equivalently, eqs.~\eqref{eqz} and \eqref{eqtau}), with initial conditions \eqref{etai} and \eqref{detai}. When $\bar k_f |\eta_f|\gg 1$, one can neglect the third and fourth terms of eq.~\eqref{eqz} with respect to $k^2$, and the solution of eq.~\eqref{eqtau} for $-1<\eta<\eta_f$ reads
\begin{equation}
\chi_k(\tau) \approx A_k e^{ik(\tau+C)}
=A_k e^{ik(\int^\eta \frac{d\bar\eta}{a(\bar\eta)}+C)}\,,
\end{equation}
where $A_k$ and $C$ are constants to be determined by the conditions at $\eta_i=-1$, and
\begin{equation}
\label{tau}
\tau=\int^\eta \frac{{\rm d}\bar\eta}{a(\bar\eta)}=
\arcsinh{[u(1+\eta)]} \,. 
\end{equation}
From $q=a^{1/2}\chi$ and conditions \eqref{etai} and \eqref{detai}, knowing that the arguments of the Hankel functions are very large under these assumptions, one gets
\begin{equation}
q_k^{(i)}(-1)=\sqrt{a_i}A_k e^{ikC}=\frac{1}{\sqrt{2\bar k_i}} e^{-i\bar k_i}\,,
\end{equation}
yielding
$C=-u$ and $A_k=1/\sqrt{2k}$. Hence,
\begin{equation}
q_k^{(i)}(\eta)=\sqrt{\frac{a(\eta)}{2k}} e^{ik\left[
 \arcsinh{u(1+\eta)}-u\right]}\,.
\end{equation}
In the region $|\eta|=|\eta_f|+\epsilon$, $|\epsilon|\ll1$ (remembering that $|\eta_f|=10^{-4}$), we obtain
\begin{equation}
\label{qietaf}
q_k^{(i)}(\eta)=
\frac{e^{i\phi(k,u)}}{\sqrt{2\bar k_f}}e^{i\bar k_f\eta \left[
1-\frac{u^2}{2(1+u^2)}\eta+O(\eta^2)
\right]}\,,
\end{equation}
with 
$\phi(k,u)=k(\arcsinh{u}-u$). 
It then follows from \eqref{qfetaf} and
\eqref{qietaf} that 
\begin{equation}
|\beta_k^{(i,f)}|^2    
=\frac{|\eta_f|^2}{4}\left[\frac{u^4}{(1+u^2)^2}+O(\eta_f)
\right]+O\left(\frac{1}{(\bar k_f\eta_f)^4}\right).
\label{betaa}
\end{equation}
Hence, for $\bar k_f|\eta_f|\gg 1$, particle production goes to a constant and very small value for any $u$. In particular, 
\bea
|\beta_k^{(i,f)}|^2     
&=& \frac{|\eta_f|^2}{4}\left[1+O(\eta_f)
\right]\,, \qquad  u\gg 1\,,  \\ \nonumber
|\beta_k^{(i,f)}|^2   
&=& \frac{|\eta_f|^2u^4}{4}\left[1+O(\eta_f)
\right]\,, \quad u \ll 1\,.
\label{betau}
\eea
Since in both cases $|\beta_k^{(i,f)}|^2$ is a small number, 
we have shown that particle production is negligible for
\be
\label{cutoff}
k\gg k_{\rm cut-off} \simeq \frac{\sqrt{1+u^2}}{u|\eta_f|}\,,
\ee
being the cut-off larger when $u$ is smaller. The estimations
following from eqs.\eqref{betaa} and \eqref{cutoff} are numerically verified in the examples presented below. \\

\paragraph{ $\bar k_f |\eta_f| \ll 1$, $0<u\ll1$\,.} From eq.~\eqref{tau} we obtain that $\tau \approx u(1+\eta) =: u + \epsilon$, where $\epsilon = u\eta$ is a small negative number as $-1 \leq \eta < 0$, and $0< u \ll 1$. In terms of the new time $\epsilon$, eq.~\eqref{eqz} reads
\begin{equation}
    \frac{{\rm d}^2\chi_k}{{\rm d}\epsilon^2}+\left[k^2-\frac{(2-m^2)}{\epsilon^2}-\frac 1 z \frac{{\rm d}^2z}{{\rm d}\epsilon^2}\right]\chi_k=0\,.
    \label{eqz-apr}
\end{equation}
In this regime, and for $k^2 \gg \frac{1}{z}\frac{d^2z}{zd\epsilon^2}$ (which is true for almost all allowed values of $k$), the general solution of eq.~\eqref{eqz-apr} is
\begin{equation}
\label{sol-apr}
\chi_k (\epsilon) = \sqrt{|\epsilon|}\left[A_1(k){\cal H}_p^{\rm(1)}(k |\epsilon|) + A_2(k){\cal H}_p^{\rm(2)}(k |\epsilon|)\right]\,.
\end{equation}
From eq.~\eqref{etai} we obtain
\begin{equation}
q_k^{(i)}(\eta=-1) = \sqrt{\frac{\pi}{2 H^{(-)}}} {\cal H}_p^{\rm(2)}({\bar{k}}_i)\,,
\end{equation}
and then we get from eq.~\eqref{sol-apr} noting that $\epsilon_i = -u$, $a_i=1/u$, and $q=a^{1/2}\chi$, and the ``in'' mode has $A_1(k)=0$, 
$A_2(k)=[\pi/(2 H^{(-)})]^{1/2}$. Hence,
\begin{equation}
\label{qin-apr}
q_k^{(i)}(\eta) = \sqrt{\frac{\pi}{2 H^{(-)}}} \sqrt{a(\eta)u|\eta|}{\cal H}_p^{\rm(2)}({\bar{k}}_i|\eta|)\,.
\end{equation}
In the same way, from eq.~\eqref{etaf} we get for the ``out'' mode around $|\eta_f|$
\begin{equation}
\label{qout-apr}
q_k^{(f)}(\eta) = \sqrt{\frac{\pi}{2 H^{(-)}}} \sqrt{u|\eta|}{\cal H}_p^{\rm(2)}({\bar{k}}_f|\eta|)\,.
\end{equation}
We want to calculate $|\beta_k|^2$ at $\eta_f$. As for $u\ll 1$, $a_f=\frac{\sqrt{1+u^2}}{u}\approx \frac{1+u^2/2}{u}= \left(1+\frac{u^2}{2}\right) a_i$, then $\bar{k}_i|\eta_f|\approx \left(1+\frac{u^2}{2}\right)\bar{k}_f|\eta_f|\ll 1$, as, by assumption, $\bar k_f |\eta_f|\ll 1$. Hence, we can expand the Hankel functions in the solutions \eqref{qin-apr} and \eqref{qout-apr} at $\eta$ around $\eta_f$ as \footnote{Note that the ratio between the next-to-leading order term and the leading order term is $(\bar k_f |\eta_f|)^2$, which is small. In the Hankel function expansion, one might also consider the next-to-leading order term ${\bar k_f}^p{|\eta_f|}^{p+1/2}$, with ratio with respect to the leading order term $(\bar k_f |\eta_f|)^{2p}$. However, as $p=(9/4 + m^2)^{1/2}$ and $0\leq m \leq 1$, then $p > 1$, and this correction is less important.}
\begin{eqnarray}
q_k^{(i)}(\eta)&=& C_1 {\bar{k}}_i^{-p}|\eta|^{-p+1/2} + C_2 {\bar{k}}_i^{-p + 2}|\eta|^{-p+5/2}  \\
 q_k^{(f)}(\eta)&=& C_1 {\bar{k}}_f^{-p}|\eta|^{-p+1/2} + C_2 {\bar{k}}_f^{-p + 2}|\eta|^{-p+5/2}\,. 
\end{eqnarray}
Inserting this expression in eq.~\eqref{eq:nk}, we get
\begin{equation}
\left|\beta_k^{(i,f)}\right|^2 \approx \frac{4|C_1|^2|C_2|^2u^4}{(\bar k_f |\eta_f|)^{4(p-1)}}\,,
\end{equation}
with $p=(9/4 + m^2)^{1/2}$. 
We shall see below that the dependence of this result with $u$, $m$, and $k_f$ (under the constraints $\bar k_f |\eta_f| \ll 1$, $0<u\ll1$)
is verified by the numerical calculations. \\ 

\begin{figure}[t!]
 \resizebox{8.8cm}{!}{\includegraphics[angle=0]{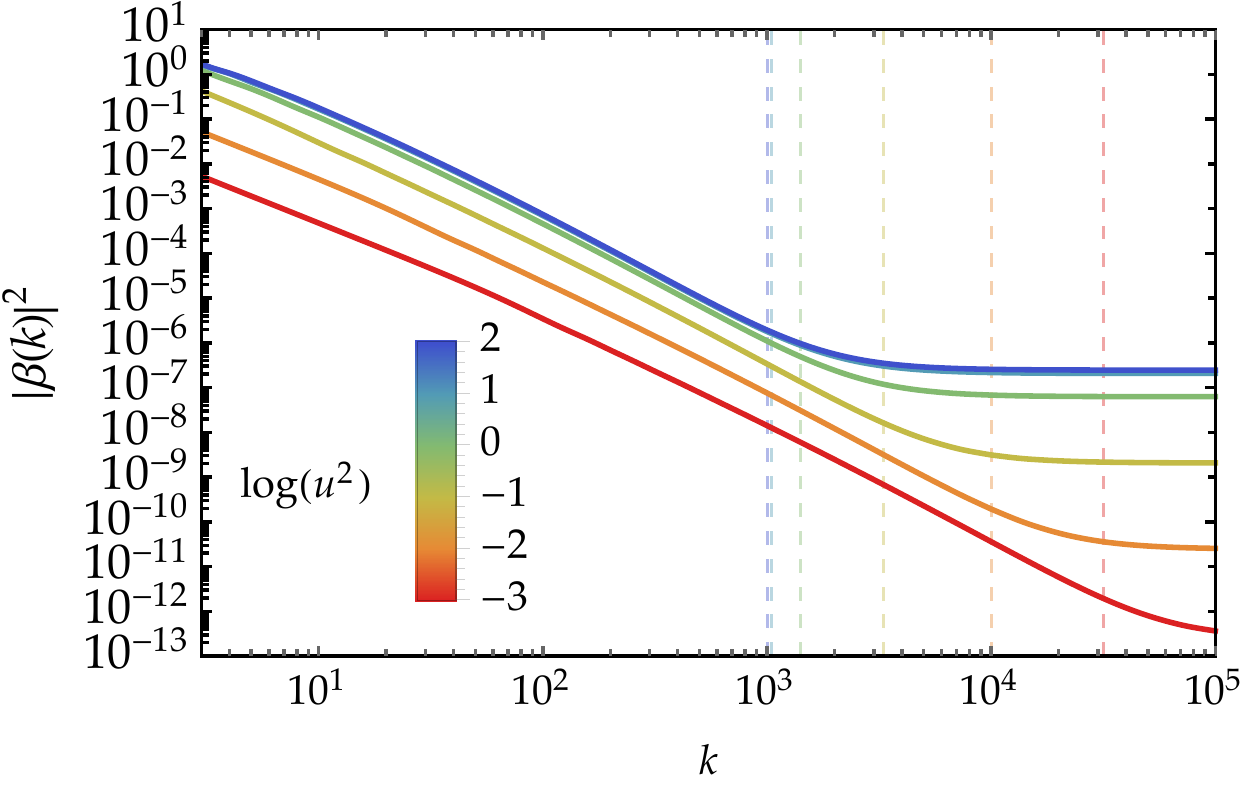}}
 \hspace{0.1cm}
  \resizebox{8.8cm}{!}{\includegraphics[angle=0]{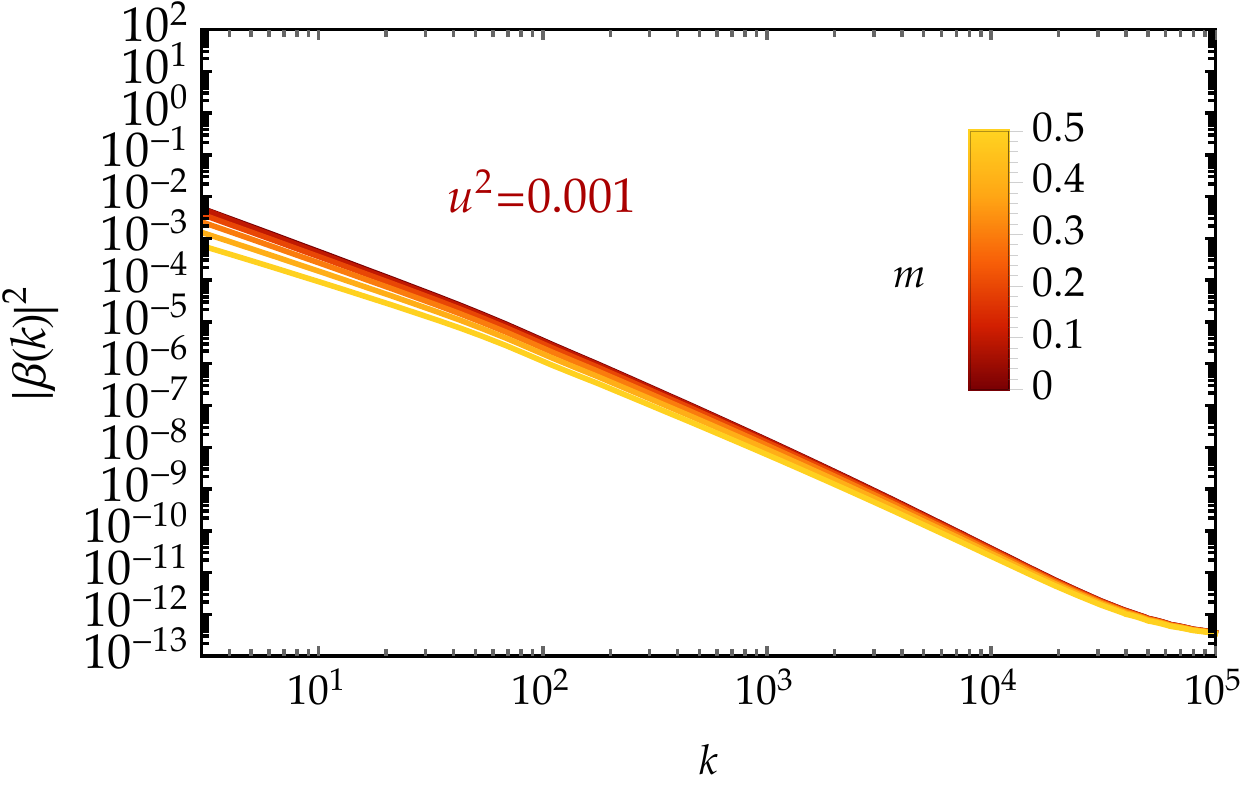}}
 \caption{The figure shows the number of particles per mode associated with the $\ket{i}$ state at the time $\eta_f$, defined in \eqref{eq:nk}. The right panel displays the massless case for the same choice of the values of $u^2$ as in the Minkowski case. In all these cases, the cut-off (dashed vertical lines) is larger when $u$ is smaller, in accordance with eq.~\eqref{cutoff}. In the right panel we show the corresponding number particle for  masses in the range $0\le m \le 0.5$ and $u^2=10^{-3}$. For fixed values $u^2$ and low frequencies the number of particles decreases when $m$ increases.}
  \label{fig:beta2dSdS}
  \end{figure}

Let us now numerically compute the number of particles given by eq.~\eqref{eq:nk} for different evolutions of the bubble, using the mode solutions obtained with eqs.~\eqref{etai} and the ``out'' vacuum in eq.~\eqref{etaf}. 
 We start by presenting the plots of $|\beta_k|^2$ as a function of $k$
for the massless case in the left panel of Fig.~\ref{fig:beta2dSdS}. They show that the creation of particles is more important for low values of $k$ and becomes negligible for $k\gg k_{\rm{cut-off}}$,  in accordance with the analytical estimates and the qualitative description presented above. The cut-off given by eq.~\eqref{cutoff} is indicated with dashed vertical lines for each value of $u^2$, after which $|\beta_k|^2$ becomes constant with $k$. 

Contrary to the case of the bubble expanding in flat spacetime discussed in the previous section,
$|\beta_k|^2$ grows with $u^2$ for any $k$ and for any $u^2$. 
This feature is related to the presence of a length scale in the dS-dS case, given by the horizon radius $R=1/H$. While bubbles with small $u^2$ have a proper radius that is larger than $R$ during their entire evolution, the radius of  bubbles with large $u^2$ starts below $R$ and reaches a value a bit larger than $R$ at the end of the evolution. Such a transition favours particle creation, as shown in the results displayed in the left panel of Fig.~\ref{fig:beta2dSdS}. 

The right panel 
of Fig.~\ref{fig:beta2dSdS} shows the results for the massive case and $u^2=0.001$ (the behaviour of $|\beta_k|^2$ for higher values of $u^2$ and different $m$ is qualitatively the same, with a smaller spread between the curves). The results show that, also in the case with curvature, $|\beta_k|^2$ decreases with increasing $m$.

 As in the Minkoswki case, continuous values of $k$ were used in the plots to optimise the display of the dependence with $u^2$, but it is worth to highlight again that $k$ can only take discrete values, corresponding to the zeros of the spherical Bessel functions. These become denser for larger values of $k$.
 
\section{Discussion}
\label{sec:discussion}

We have analysed the influence of two types of expanding bubbles, modelled as  thin walls, on the creation of particles of a test scalar field inside them.
 The squared modulus of the Bogoliubov coefficient $\beta_k$ 
was numerically computed for several trajectories of both types of bubbles, described in a unified way. The field modes were obtained under the assumption of the validity of the adiabatic approximation, in which the effects of the wall of the bubble 
are considered only through the time-dependent eigenfrequencies $\omega_{ln'}(\eta)= j_{l,n'} /a(\eta)$, which naturally appear as a result of the time dependence of the radius of the sphere. We have checked that the necessary conditions for this approximation to be valid are satisfied in both cases. From a quantitative point of view, the creation of particles is  typically 5-6 orders of magnitude larger in the dS-dS case, regardless of the value of $u^2$ and of the mass. 
 This is likely due to the nonzero curvature, the latter being absent in the Minkowski case. 

While in the case of a bubble expanding in a Minkowski background the creation of particles is influenced 
only by the intrinsic features of the bubble, represented by parameters such as its initial and final radius, and the associated velocity in the intermediate region, 
in the configuration of a bubble separating two de Sitter geometries, both those features and gravitational effects are expected to contribute to particle creation. 
Our results show that, in the adiabatic approximation, in both cases particle production is not very sensitive to the value of $m$, and is always smaller for larger masses. 
In the case of a bubble expanding in Minkowski spacetime,
the number of particles grows with $u^2$ for small $u^2$, has a maximum at a certain value of $k$ for each $u^2$, and decreases with $u^2$ for large $u^2$, until it stabilises. This may be understood as the result of the combined effects of the size of the bubble (dominant for small $u^2$) and its evolution (dominant for large $u^2$).

In the dS-dS case, 
the number of particles always increases with $u^2$, where larger $u^2$ means smaller  initial and final radius and larger difference between them (hence, larger velocity for the bubble while it is joining the initial and final states).  
In fact, it is the presence of a length scale (the horizon radius $R$) in the dS-dS case that produces a difference in particle production, which is larger for bubbles with initial radius smaller than $R$, and final radius larger than $R$.

It remains to be seen whether these results are still valid in more complex settings. In particular, a more detailed model would entail a scalar field that is not a test field, but generates the inflationary expansion inside the bubble. The fluctuations of such a scalar field will be coupled to those of the metric, and their power spectrum would display the influence of the motion of the bubble.  

Another possible extension of this work would be to consider the influence of the corrections to the thin-shell approximation (following the developments presented in
\cite{Garfinkle1989, Carter1994}) on the creation of particles. We hope to return to these issues in a future publication.


\appendix
\section{Back-reaction effects}
We shall show here that the energy of the particles produced in the dS-dS case is negligible compared to that of the background. The latter can be calculated as
\begin{equation}
E_\Lambda = \int_{V_f}d^3x\sqrt{-\gamma}\;T_0^{\;0}(\Lambda)\,,
\end{equation}
where $\gamma$ is the determinant of the metric of the 3-space.
In dimensionless form (with $c= \hbar=1$),
\begin{equation}
\bar E_\Lambda = \frac 3 2 \left(\frac{\bar a_f}{\bar\eta_f}\right)^3\frac{R^2}{G}\,.
\end{equation}
Using the values $\bar a_f\approx 1$ (see right panel of fig.~\ref{bubevol}), $\bar\eta_f\approx10^{-4}$, and $H=10^9$ GeV (with $R=1/H$), it follows that 
$\bar E_\Lambda \approx 10^{32}$.

The energy of the created particles is given by
\begin{equation}
\bar E = \sum_{l,n}n^{\rm (i,f)}_{ln}
\bar\nu_{ln}^{\rm (f)}\,, 
\label{energy}
\end{equation}
with
$\bar\nu^{\rm (f)}_k\equiv \sqrt{(k/\bar a_{\rm (f)})^2+\bar m^2}$, and $k\equiv j_{l,n}$.
In order to obtain an upper bound for $\bar E$, we shall consider only the case $m=0$ (for which the creation of particles is more important, see bottom panels in  Fig.~\ref{fig:beta2dSdS}). Hence,
\begin{equation}
\bar E < \frac{n_M}{\bar a_{\rm (f)}} \sum_{l,n} j_{l,n}\,,
\label{energy2}
\end{equation}
where $n_M$ is the maximum value of $n^{(i,f)}_{ln}$.
In order to establish the largest value of $l$ that contributes to the sum in eq.~\eqref{energy},
we shall use the fact that the zeros of the Bessel functions obey the so-called interlacing property
\cite{watson}:
\begin{equation}
0<j_{l,1}<j_{l+1,1}<j_{l,2}<j_{l+1,2}<j_{l,3}<\ldots 
\end{equation}
As seen in fig.~\ref{fig:beta2dSdS}, there is a cut-off for the particle production, which will be taken as 
$k_M=10^4$. Hence, the maximum $l$, denoted by $l_M$,
is set by $j_{l_M,1}=k_M$. Since for large $l$ we have that 
$j_{l,1}\approx l + O(l^{1/3})$
\cite{stegun}, we can adopt $l_M=10^4$.  Hence, an upper limit for $\bar E$ is given by
\begin{equation}
\bar E < 10^5 \sum_{n=0}^N j_{0,n}\,,
\label{energy3}
\end{equation}
where $N$ is determined by $j_{0,N}\approx 10^4$, yielding $N\approx 3200$.
The upper limit defined by eq.~\eqref{energy3} is clearly an overestimation, but is nevertheless useful for our purposes.  With the help of numerical calculation, we obtain
$\bar E \lesssim 10^{12}$. Hence, $\bar E / \bar E_\Lambda \lesssim 10^{-20}$.

\begin{acknowledgments}
FATP would like to thank Jose Beltrán Jiménez for very useful comments and discussions. FATP acknowledges support from the Atracci\'on del Talento Cient\'ifico en Salamanca programme,
from project PGC2018-096038-B-I00 by Spanish Ministerio de Ciencia, Innovaci\'on y Universidades and Ayudas del Programa XIII by USAL,  and the hospitality of UERJ and CBPF. This work was started while FATP was benefiting from a CAPES National Brazilian Fellowship at UERJ. SEPB acknowledges support from UERJ. NPN acknowledges the support of CNPq of Brazil under grant PQ-IB
309073/2017-0.
\end{acknowledgments}

\bibliographystyle{apsrev} 
\bibliography{bibliography} 

\end{document}